\documentclass{article}
\usepackage{spconf,amsmath,graphicx}
\usepackage{amssymb}
\usepackage{url}
\usepackage{booktabs}
\usepackage{enumitem}
\setlist{nosep, leftmargin=14pt}

\usepackage{mwe} 

\title{Spatial-Temporal Convolutional Attention for Mapping Functional Brain Networks}

\name{Yiheng Liu$^1$, Enjie Ge$^1$, Ning Qiang$^1$, Tianming Liu$^2$ and Bao Ge$^1$}
\address{${}^1$School of Physics \& Information Technology, Shaanxi Normal University, Xi'an, China\\
${}^2$Department of Computer Science, University of Georgia, GA, USA}

\begin{document}
%
\maketitle
%
\begin{abstract}
Using functional magnetic resonance imaging (fMRI) and deep learning to explore functional brain networks (FBNs) has attracted many researchers. However, most of these studies are still based on the temporal correlation between the sources and voxel signals, and lack of researches on the dynamics of brain function. Due to the widespread local correlations in the volumes, FBNs can be generated directly in the spatial domain in a self-supervised manner by using spatial-wise attention (SA), and the resulting FBNs has a higher spatial similarity with templates compared to the classical method. Therefore, we proposed a novel Spatial-Temporal Convolutional Attention (STCA) model to discover the dynamic FBNs by using the sliding windows. To validate the performance of the proposed method, we evaluate the approach on HCP-rest dataset. The result indicate that STCA can be used to discover FBNs in a dynamic way which provide a novel approach to better understand human brain.
\end{abstract}
\begin{keywords}
fMRI, Attention Mechanism, Functional Brain Network, Brain Function Dynamic
\end{keywords}
\section{Introduction}
\label{sec:intro}
Functional magnetic resonance imaging (fMRI) is a non-invasive imaging method to estimate brain function which indirectly detect the activities of brain function by measuring the blood-oxygen level dependence (BOLD) \cite{fmri,fmri2}. To explore the functional brain networks (FBNs), various analytical approaches have been proposed, each providing a different model for mapping brain functional pattern. Most of the previous methods for mapping FBNs using fMRI are based on the temporal correlation between the source and voxel signals, such as indenpendent component analysis (ICA) \cite{7} and sparse dictionary learning (SDL) \cite{10}. Many of these approaches can be viewed as a blind source separation problem, the fMRI signals are modeled as a linear combination of the sources and the temporal patterns \cite{space}. Although significant results have been achieved by these methods, the process of extracting the sources is still in a linearity way, which makes these classical methods limited by their shallow nature. 

Recently, to overcome the shallow nature of the linear models, various of deep learning based methods have been proposed to discover the FBNs. Most of these methods are based on the autoencoders, they use different autoencoders to extract the sources in an self-supervised manner, and then use the generative linear model, such as LASSO to generate the FBNs \cite{MAMSM,zhao2022embedding}. In general, these deep learning based methods can indeed extract better encoder representations as the sources than the classical methods, such as ICA and SDL, but still generate FBNs in a linear and independent manner, with the sources extraction and the FBNs generation as 2 separate steps. Generating the FBNs in such way is time-consuming and does not fully utilize the advantages of deep learning, and cannot directly generate the FBNs with deep learning. In fact, the FBNs should be able to be represented at the feature extraction stage if we directly extract the spatial features, and generating the FBNs in a linear and independent way would require more longer source signals to increase confidence. At the same time, there is still a lack of methods for using deep learning to generate the dynamic FBNs. The mainstream methods usually adopt sliding window technology to generate the dynamic FBNs, they usually set the window size and the sliding step, and then input the data into commonly used classical models, such as ICA \cite{kiviniemi2011sliding} and SDL \cite{DFBNSDL}. In addition to sliding window approaches, there are also some methods tried to discover the spatial pattern in dynamic brain function, such as co-activation patterns (CAPs) \cite{CAP1,CAP2,CAP3}. The most widely used CAP model assumes that one CAP is active at each time point, and each CAP represents a spatial pattern of the source, usually clustering the CAP based on spatial similarity. Such a view oversimplified about the brain function and can only provides a limited amount of CAP.

To solve these problems, we proposed a novel deep learning based method called Spatial-Temporal Convolutional Attention (STCA) to directly generate the high quality dynamic FBNs by the sliding window technology. The STCA model is based on the convolutional neural network (CNN) architecture and trained in a self-supervised manner. The CNN architecture can make full use of the FBN's prior knowledge which is the widespread local correlation in the fMRI signals, it is suitable for modeling the fMRI signals naturally. To validate the performance of the STCA model, we perform the experiments on the HCP-rest dataset \cite{hcp} which is a resting-state fMRI (rsfMRI) dataset. The result demonstrate that STCA can generate dynamic FBNs more better than the mainstream methods, such as ICA and SDL, the generated FBNs have higher spatial similarity to the templates. Source code of this work can be found in \url{https://github.com/SNNUBIAI/STCAE}.

\section{Methods}

\subsection{Dataset and preprocessing}
\label{sec:Dataset}
In this paper, we adopt the Human Connectome Project (HCP) 900 Subjects Data Release \cite{hcp} as the experiments data. The HCP-rest is a publicly accessible rsfMRI dataset (available on \url{https://db.humanconnectome.org}.). Each individual contains 1200 time points, all of these data have been officially preprocessed and registered to the MNI152 $4\times4\times4$ standard template space. We masked the fMRI data and resampled it to a size of $49\times58\times47$.

\subsection{Spatial-Temporal Convolutional Attention (STCA)}
\label{sec:STCA}
\begin{figure*}[h]
	\centerline{\includegraphics[width=\textwidth]{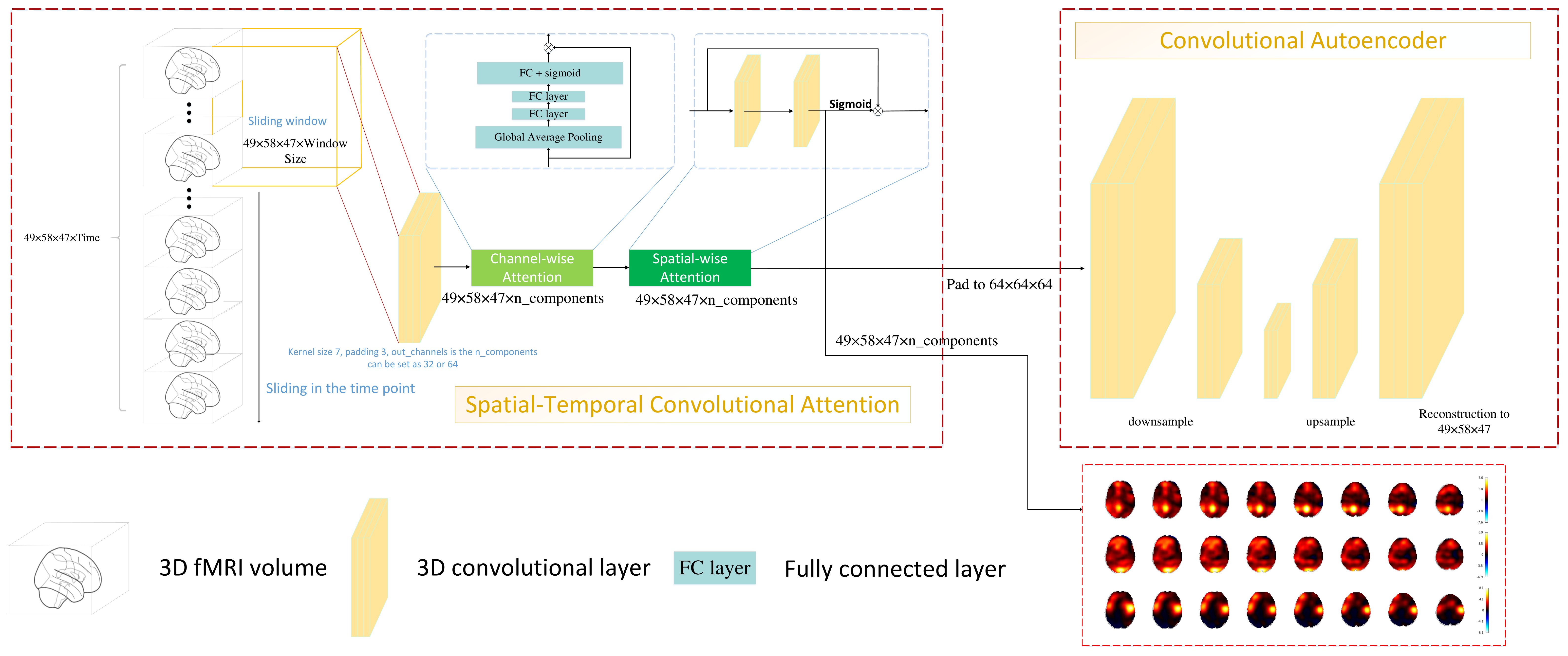}}
	\caption{The architecture of the Spatial-Temporal Convolutional Attention. The STCA used the spatial and channel-wise attention in the spatial and temporal of the fMRI signals to discover the FBNs. The convolutional autoencoder module is used to guide the attention modules to focus on the activated regions in the training stage, it is not needed in the evaluate stage. The spatial-wise attention module's weight map are transformed to z-score values, which represent the degree of functional activation on space, that is, the FBNs are derived. The results are close to the known FBNs derived by the classical methods, such as ICA and SDL.}
	\label{fig:STCA}
\end{figure*}

The Fig~\ref{fig:STCA} summarizes the proposed Spatial-Temporal Convolutional Attention (STCA) based pipeline for modeling the FBNs. 
Before going into the architectural details, we will briefly explain the design idea. Since the existing mainstream practice is to extract the sources through the encoder and then generate the FBNs with a generative linear model. Extracting sources and constructing FBN are two separate steps, the noise is unavoidable in the process of generating the FBNs, and usually need to take longer time points of the fMRI signals to ensure the confidence. If the deep neural network can extract the sources, the spatial pattern of the sources should be represented in the feature extraction step. In fact, the spatial pattern of the sources or called the FBNs, have a local correlation in the spatial and temporal pattern. The local correlation can be easily observed in the activation maps shown in Fig~\ref{fig:compare}, the 10 resting state network (RSN) derived by ICA \cite{ICARSNTemplate}. The activated voxels tend to cluster together in the spatial and the voxels in the activation regions are also correlated with each other in the temporal. These spatial patterns widely exists in the brain functional activities, inspired by the application of the spatial-wise attention (SA) in computer vision tasks \cite{CBAM}, we tried to introduce the SA to generate the FBNs, tried to give higher weights to the activation regions. So that the output of the SA module can be define as the FBNs.

The goal of our task is to train a neural network: 
\begin{equation}
	f:X_{dyn} \longrightarrow S_{dyn}
\end{equation}
where $X_{dyn} = (X_{1}, ..., X_{t})$ is the sequence of the fMRI volumes with $t$ time points and $S_{dyn} \in \mathbb{R}^{N \times D \times H \times W}$ is the weight maps of the volumes $X_{dyn}$ with number of the weight maps $N$, the depth of the weight map $D$, the height of the weight map $H$ and the width of the weight map $W$. The STCA is an independent attention module, which cannot directly generate the weight maps in the absence of the labeled data. It needs to be trained in a self-supervised manner to guide the attention module to learn the FBNs. Autoencoders are commonly used self-supervised training methods, which compress the high-dimensional information to low-dimensional information to extract the sources. Guided by the autoencoder, the STCA learns to focus on the spatial patterns of the source to help the encoder generate better encoder representations as the sources, so that the spatial pattern of the sources can be represented at the feature extraction stage. 

The STCA consists of a 3D convolutional layer, a channel-wise attention (CA) module and a spatial-wise attention (SA) module shown in Fig~\ref{fig:STCA}. The convolutional autoencoder is just to guide the SA module and does not involve in the evaluate stage. The convolutional layer and CA module are both designed to help extract more useful features, only the SA module is used to locate the regions where the FBNs are. We represent the $GAP$ as global average pooling which will transform the 3D volumes to 1D vector, $W$ as the fully connected layer,  $sigmoid$ as the activation function that can map the features to $[0, 1]$. The CA can be represented as:
\begin{equation}
	CA = sigmoid(\sigma(\sigma(GAP(f^{l}){W_{1}}){W_{2}}){W_{3}})
\end{equation}
\begin{equation}
	\widetilde{f}^{l} = f^{l} \circ CA 
\end{equation}
We represent the 3D convolutional layer's output features as $f^{l} \in \mathbb{R}^{C\times D\times W\times H}$, $conv3D$ as the 3D convolutional layer, $\sigma$ as the activation function $GELU$, $\widetilde{f}^{l}$ as final output of the SA module. The SA can be represented as:
\begin{equation}
	SA = conv3D(\sigma(conv3D(f^{l}))) 
\end{equation}
\begin{equation}
	\widetilde{f}^{l} = f^{l} \circ sigmoid(SA) 
\end{equation}

\subsection{Generation FBNs based on the STCA}
\label{sec:fbn}
We adopt the sliding window technique to discover the dynamic FBNs. Here, the window size is set as 40, stride is set as 1. Like the classical method, such as ICA and SDL, both training and testing are performed in the same individual. When training the STCA, the loss function of the autoencoder is mean square error (MSE), the learning rate is set as 0.0001, the optimizer is Adam and the epoch is set as 1. We convert $SA$ to z-score, and the threshold is set as 3.6.
\section{Experimental Results}
\subsection{Overview of the weight maps derived by STCA}
\begin{figure}[h]
	\centerline{\includegraphics[width=\columnwidth]{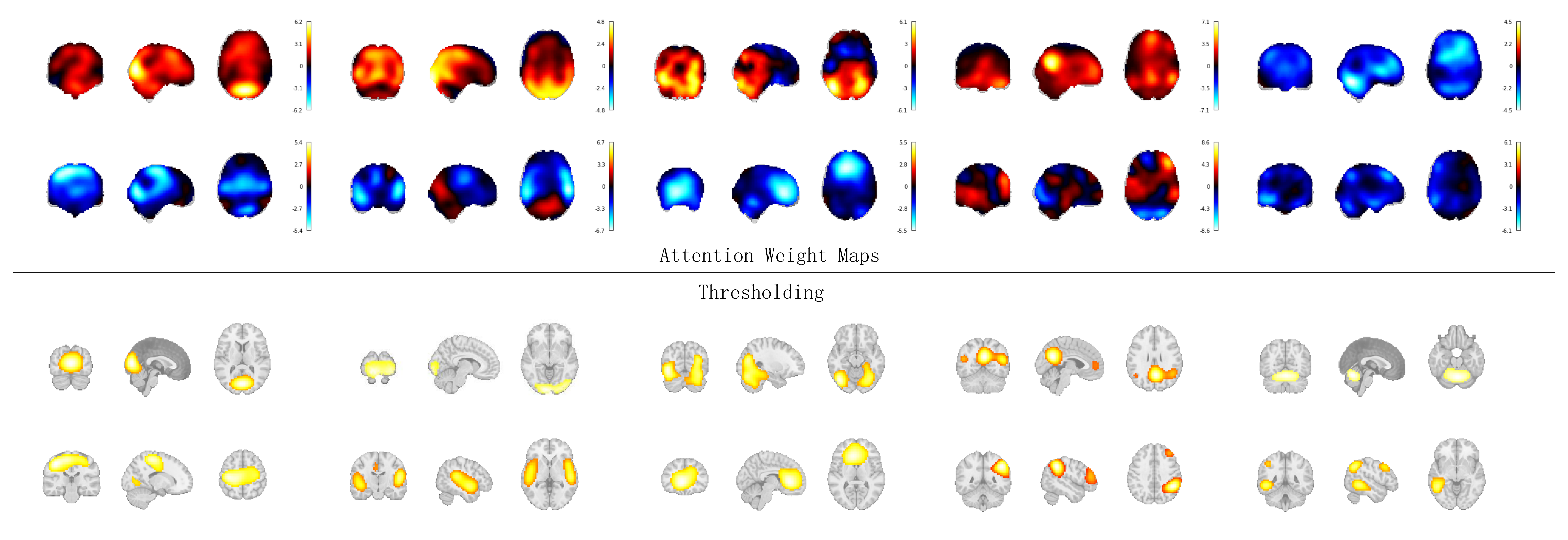}}
	\caption{The weight and thresholded maps derived by STCA.}
	\label{fig:weight_map}
\end{figure}

The representative weight maps derived from the STCA are shown in the upper panel of the Fig~\ref{fig:weight_map}, these weight maps are all convert into z-score, the thresholded maps shown in the lower panel. It can be seen that regions activated with high weights can be easily explained by the known functional networks, such as visual network, default mode network, etc. This result demonstrate that the spatial pattern of the sources can indeed be captured by the spatial-wise attention mechanism in the feature extraction stage which means that we provide a new paradigm for discovering the functional networks.

To evaluate the quality of the FBNs generated by STCA, we calculated the spatial similarity between the generated FBNs and ten resting-state templates and compared the results with ICA and SDL. The spatial similarity is defined by the intersection over union rate (IoU) between two FBNs $N^{(1)}$ and $N^{(2)}$, where $n$ refers to the volume size:
\begin{equation}
	IoU(N^{(1)}, N^{(2)}) = \frac{\sum_{i=1}^{n}|N_{i}^{(1)} \cap N_{i}^{(2)}|}{\sum_{i=1}^{n}|N_{i}^{(1)} \cup N_{i}^{(2)}|}
\end{equation}

\begin{figure}[h]
	\centerline{\includegraphics[width=\columnwidth]{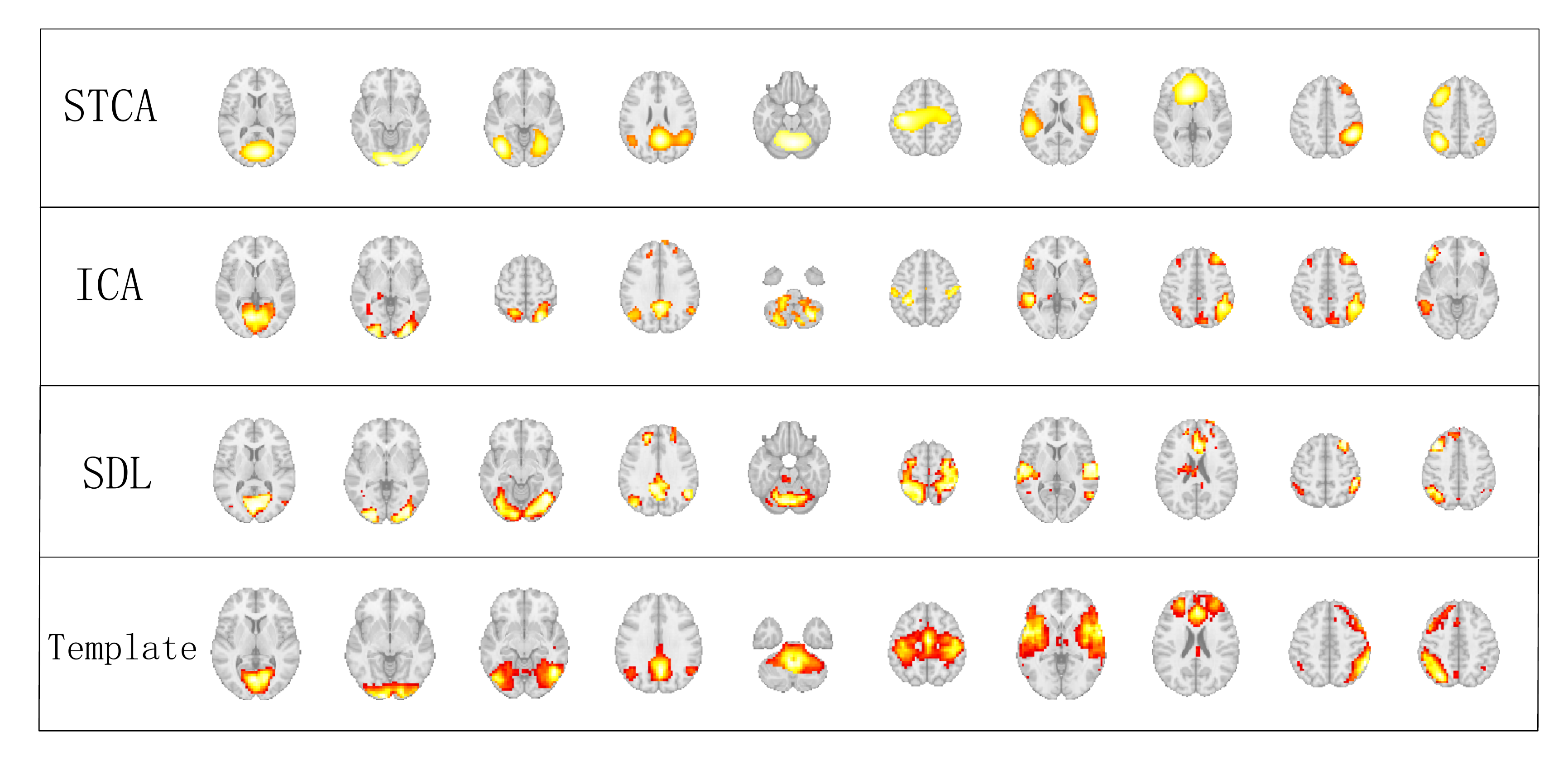}}
	\caption{Comparision with the classical methods, such as ICA and SDL. The templates are based on \cite{ICARSNTemplate}}
	\label{fig:compare}
\end{figure}

The similarity metrics $IoU(N_{STCA}, N_{template})$, $IoU($ $N_{SDL}, N_{template})$ and $IoU(N_{ICA}, N_{template})$ are used to evaluate the performance of STCA. We randomly selected an individual to do the experiments, the results of the evaluation are shown in Fig~\ref{fig:compare} and Table~\ref{tab:iou_template}, the 10 RSN templates are in line 4. The results indicate that the proposed STCA can generate better FBNs than ICA.

\begin{table}[h]
	\caption{IoU with templates in an individual in HCP-rest dataset.}\label{tab:iou_template}
	\centering
	\begin{tabular}{cccc}
		\toprule
		Template & ICA & SDL & STCA\\
		\midrule	
		RSN 1  & 0.6273 & 0.6053 & 0.6505\\	
		RSN 2  & 0.4260 & 0.4554 & 0.5695\\	
		RSN 3  & 0.2414 & 0.3340 & 0.3446\\	
		RSN 4  & 0.2889 & 0.3446 & 0.4250\\	
		RSN 5  & 0.1932 & 0.2635 & 0.4474\\	
		RSN 6  & 0.0888 & 0.2279 & 0.3900\\	
		RSN 7  & 0.1179 & 0.1992 & 0.4284\\	
		RSN 8  & 0.0540 & 0.1951 & 0.3159\\	
		RSN 9  & 0.1843 & 0.1890 & 0.2628\\	
		RSN 10  & 0.2436 & 0.1920 & 0.2597\\
		\bottomrule
	\end{tabular}
\end{table}
\subsection{Dynamic FBNs in resting state fMRI}
\begin{figure}[h]
	\centerline{\includegraphics[width=\columnwidth]{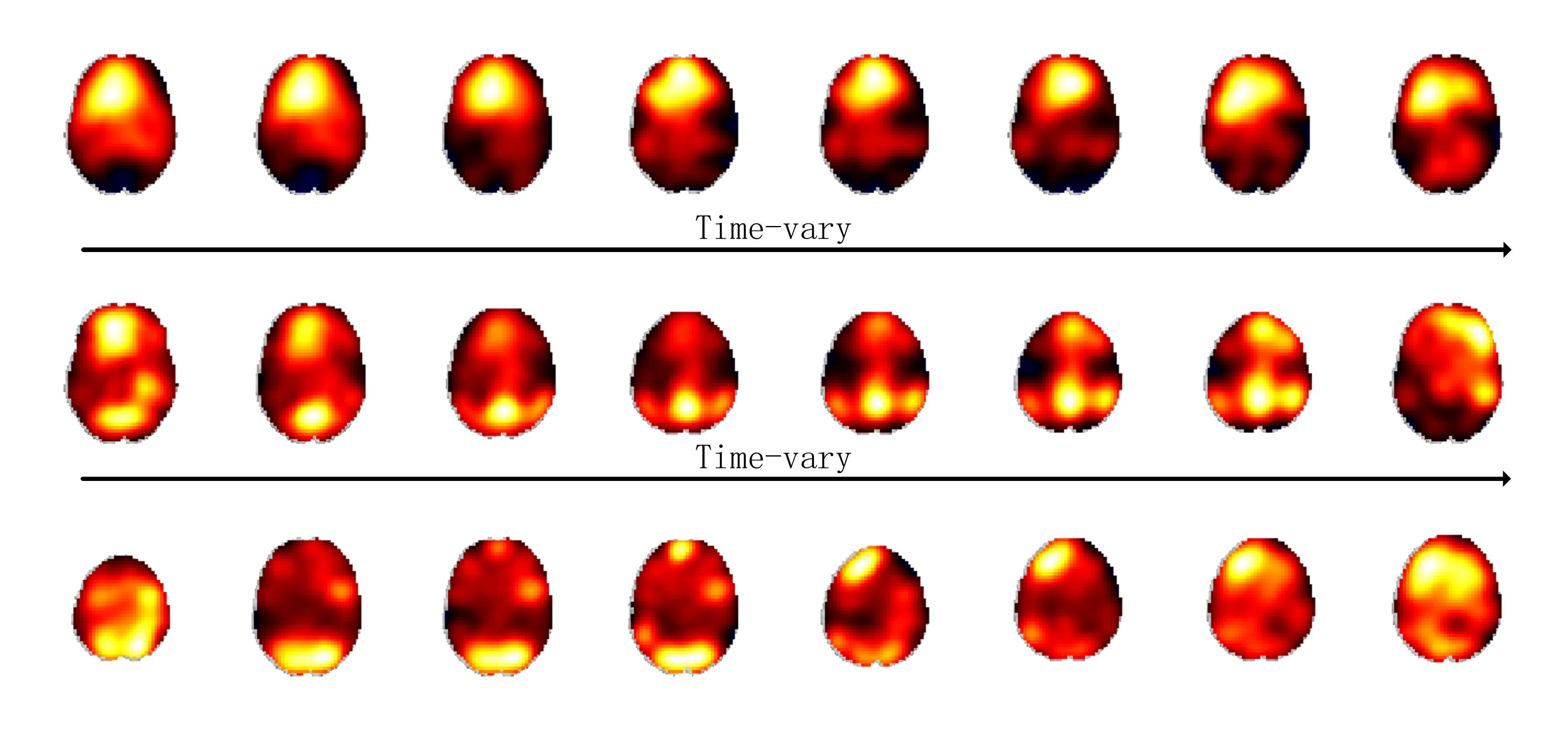}}
	\caption{Some representative FBN's dynamic change process.}
	\label{fig:dfbn}
\end{figure}

\begin{figure}[h]
	\centerline{\includegraphics[width=\columnwidth]{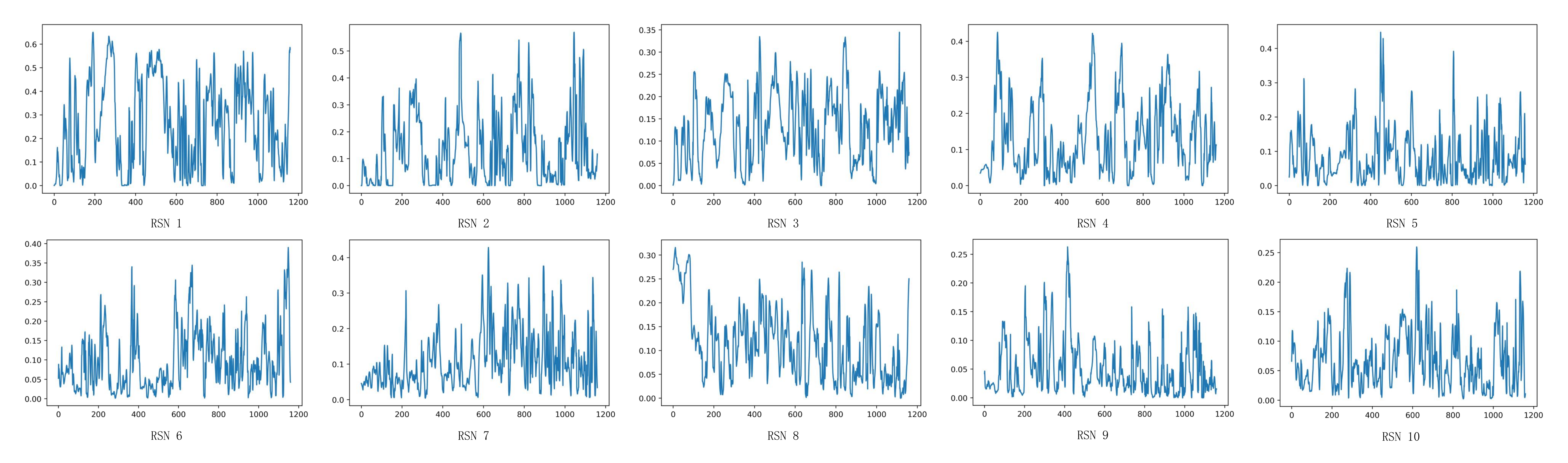}}
	\caption{The spatial similarity change between the FBNs generated by each window and 10 RSN templates over time. The horizontal axis is the change of the time window, and the vertical axis is the IoU with the RSN template.}
	\label{fig:dynamic}
\end{figure}

The proposed STCA can be combined with sliding window technique to discover dynamic FBNs. Here, the window size is set as 40, stride is set as 1. We plot some representative functional network's dynamic change process shown in Fig~\ref{fig:dfbn}, which shows that some FBNs' transferring processes. We can easily find that the RSN fade in or fade out over time.Quantitatively, we also plot the spatial similarity change between the FBNs generated by each window and 10 RSN templates over time, which are shown in Fig~\ref{fig:dynamic}, we can also see the fade in and fade out process in form of IoU values.
\section{Conclusion}
The proposed STCA provides a new idea and framework to construct the FBNs via the spatial-wise attention, and shows excellent performance on generating the FBNs. It also provides a new tool for studying dynamic FBNs. In the future, we plan to study brain spatial state transition based on it and perform new experiments on task-based fMRI (tfMRI). We also plan to perform the experiments on functional connectivity (FC) analysis and diseases classification based on the proposed STCA. 
\section{Acknowledgments}
\label{sec:acknowledgments}
The work was supported by the National Natural Science Foundation of China (NSFC61976131).

\bibliographystyle{IEEEbib}
\bibliography{ref}

\end{document}